\documentclass[journal]{IEEEtran}
\usepackage[utf8]{inputenc}
\usepackage{amsmath}
\usepackage{braket}
\usepackage[pdftex]{graphicx}
\usepackage{amsfonts}
\usepackage{pgfplots}
\usepackage{csquotes}
\usepackage{dsfont}
\usepackage{hhline}
\usepackage{amssymb}
\usepackage{listings}
\usepackage{color}
\usepackage{cite}
\usepackage{array}
\usepackage{hyperref}
\definecolor{codegreen}{rgb}{0,0.6,0}
\definecolor{codegray}{rgb}{0.5,0.5,0.5}
\definecolor{codepurple}{rgb}{0.58,0,0.82}
\definecolor{backcolour}{rgb}{0.95,0.95,0.92}
 
\lstdefinestyle{mystyle}{
    backgroundcolor=\color{backcolour},   
    commentstyle=\color{codegreen},
    keywordstyle=\color{magenta},
    numberstyle=\tiny\color{codegray},
    stringstyle=\color{codepurple},
    basicstyle=\footnotesize,
    breakatwhitespace=false,         
    breaklines=true,                 
    captionpos=b,                    
    keepspaces=true,                 
    numbers=left,                    
    numbersep=5pt,                  
    showspaces=false,                
    showstringspaces=false,
    showtabs=false,                  
    tabsize=2
}
 
\usepackage{subfigure}

\usepackage{dcolumn}
\usepackage{tabularx}
\usepackage{longtable}
\usepackage{braket}
\usepackage{float}
\newcolumntype{C}{>{\centering\arraybackslash}X}
\renewcommand{\arraystretch}{1.5}

\begin{document}

\title{Measurement-device-independent QSDC protocol using Bell and GHZ states on quantum simulator}

\author{Arunaday~Gupta, Bikash~K.~Behera and~Prasanta~K.~Panigrahi}
\maketitle
\begin{abstract}
Secure cryptographic protocols are indispensable for modern communication systems. It is realized through an encryption process in cryptography. In quantum cryptography, Quantum Key Distribution (QKD) is a widely popular quantum communication scheme that enables two parties to establish a shared secret key that can be used to encrypt and decrypt messages. But security loopholes still exist in this cryptographic protocol, as an eavesdropper can in principle still intercept all the ciphertext to perform cryptanalysis and the key may get leaked to the eavesdropper, although it happens very rarely. However, there exists a more secure quantum cryptographic scheme known as Quantum Secure Direct Communication (QSDC) protocol that eliminates the necessity of key, encryption and ciphertext transmission. It is a unique quantum communication scheme where secret information is transmitted directly over a quantum communication channel. We make use of measurement-device-independent (MDI) protocol in this scheme where all the measurements of quantum states during communication are performed by a third party that can be untrusted or even an eavesdropper. This eliminates all loopholes in practical measurement devices. Here, we realize this MDI-QSDC protocol using Bell and GHZ states in the IBM Quantum Experience platform and implement swapping circuits for security check.
\end{abstract}


\section{Introduction}
In order to ensure secure communication, various quantum cryptographic schemes have been developed that makes it rather impossible for the information to leak out to an eavesdropper. During a communication process between two parties that involves encryption and decryption process, a key is required. Shor's algorithm \cite{qsdc_Shor1994} makes it possible for an eavesdropper to steal the key. In order to overcome this, a very popular quantum cryptographic scheme, Quantum Key Distribution (QKD) is used which enables both the parties to establish a shared secret key. Since Bennet and Brassard introduced the BB84 protocol \cite{qsdc_Bennett1984}, a number of QKD protocols have been developed, most notably the Ekert protocol\cite{qsdc_Ekert1991}. However, there are still other security loopholes in the QKD scheme. The eavesdropper can still intercept all the ciphertext and perform cryptanalysis. Although rare, the secret information may leak to the eavesdropper. 

However, another ingenious quantum cryptographic scheme known as the Quantum Secure Direct Communication (QSDC) protocol has been developed and several theoretical and experimental works \cite{qsdc_Niu2018,qsdc_Deng2003,qsdc_Zheng2014,qsdc_Hu2016,qsdc_Zhang2017,qsdc_Zhu2017,qsdc_JLi2015} have been carried out on the same. This scheme does not require key and encryption. It is a novel secure communication system, without key distribution, key storage and management, and ciphertext. This can offer significant advantage in terms of security of communication, and provides a great new alternative in the field of cryptographic technology.

The notable distinction between the QSDC and QKD protocol is that the security is known only after the key distribution session is completed in QKD, whereas QSDC first establishes the security of the quantum channel. In practical quantum communication systems, defects in measurement devices can lead to leakage of secret information to an eavesdropper without even being detected. 

We can fix this problem using measurement-device-independent (MDI) protocol. MDI-QKD \cite{qsdc_Vazirani2019,qsdc_Lo2012} and MDI-QSDC \cite{qsdc_Niu2018} protocols have been developed and here we realize the MDI-QSDC protocol using Bell and GHZ states in the IBM Quantum Experience (IBM QE) platform to implement the quantum circuits and compute the results. IBM QE, an open-access platform for giving access to quantum simulators and real chips, has been widely used recently to execute several research works in the field of quantum simulation among which, demonstration of path integral formalism \cite{qsdc_Gupta2020}, quantum harmonic oscillator \cite{qsdc_BaishyaarXiv2019,qsdc_JainaRG2019}, quantum gravity \cite{qsdc_ManabQIP2020}, Klein-Gordon equation \cite{qsdc_KapilarXiv2018}, quantum tunneling \cite{qsdc_HegadearXiv2018}, observation of Berry phase \cite{qsdc_MalikarXiv2019}, variational quantum eigensolver \cite{qsdc_KumararXiv2019} are a few of them.

In the MDI-QSDC protocol, during the security check, we make use of the swapping circuit to calculate the inner product of two-qubit states that we developed in our previous paper \cite{qsdc_Gupta2020} to detect any act of eavesdropping.

\section{MDI-QSDC protocol using Bell states \label{qsdc_Sec2}}

We have two communicating parties, Alice and Bob, both randomly prepare Bell states and an untrusted third party Charlie who performs measurements. For our convenience here, the Bell states prepared are any one of the following-

\begin{eqnarray}
\ket{\psi^{+}}=\frac{1}{\sqrt{2}}(\ket{00}+\ket{11})\\
\ket{\psi^{-}}=\frac{1}{\sqrt{2}}(\ket{00}-\ket{11})
\end{eqnarray}\\\\

For both Alice and Bob, we prepare four Bell states as follows-

\setlength{\arrayrulewidth}{0.5mm}
\setlength{\tabcolsep}{30pt}
\renewcommand{\arraystretch}{3.5}
\begin{table}[H]
\caption{Preparation of Bell states}
\begin{tabular}{|c|c|c|}
\hline
Position & Alice               & Bob \\ \hline
1        & $\ket{\psi^{+}}$    & $\ket{\psi^{-}}$  \\
2        & $\ket{\psi^{+}}$    & $\ket{\psi^{+}}$  \\
3        & $\ket{\psi^{-}}$    & $\ket{\psi^{-}}$  \\
4        & $\ket{\psi^{+}}$    & $\ket{\psi^{-}}$  \\ \hline
\end{tabular}
\label{table:1}
\end{table}

Now, we insert 1-qubit states of computational and Hadamard basis at random positions of Alice and Bob-

\setlength{\arrayrulewidth}{0.5mm}
\setlength{\tabcolsep}{30pt}
\renewcommand{\arraystretch}{3.5}
\begin{table}[H]
\caption{Random insertion of 1-qubit states}
\begin{tabular}{|c|c|c|}
\hline
Position &Alice            & Bob \\ \hline
1        &$\ket{\psi^{+}}$ &$\ket{\psi^{-}}$ \\
2        &$\ket{+}$        &$\ket{-}$\\
3        &$\ket{\psi^{+}}$ &$\ket{\psi^{+}}$ \\
4        &$\ket{\psi^{-}}$ &$\ket{\psi^{-}}$ \\
5        &$\ket{0}$        &$\ket{\psi^{-}}$ \\ 
6        &$\ket{\psi^{+}}$ &$\ket{1}$\\ \hline
\end{tabular}
\label{table:2}
\end{table}
We assign one of the qubits from the Bell states at different positions of Alice and Bob respectively to each of them denoted by A and B respectively.

The other qubits of the pairs, along with the randomly inserted 1-qubit states denoted by C and D respectively are sent to Charlie, the third party in this protocol.

Bell basis measurements are now performed on the qubits sent to Charlie. Charlie announces the results of the Bell basis measurements. Due to entanglement swapping \cite{qsdc_Pathak2013}, the qubits in A and B become entangled-

\begin{equation}\label{qsdc_eq3}
    \begin{split}
        \ket{\psi^{+}}_{AC}\otimes\ket{\psi^{+}}_{BD}=\frac{1}{2}\big[\ket{\psi^{+}}_{AB}\ket{\psi^{+}}_{CD}\\+\ket{\psi^{-}}_{AB}\ket{\psi^{-}}_{CD}\\+\ket{\phi^{+}}_{AC}\ket{\phi^{+}}_{BD}\\+\ket{\phi^{-}}_{AC}\ket{\phi^{-}}_{BD}\big]\\
        \ket{\psi^{+}}_{AC}\otimes\ket{\psi^{-}}_{BD}=\frac{1}{2}\big[\ket{\psi^{+}}_{AB}\ket{\psi^{-}}_{CD}\\+\ket{\psi^{-}}_{AB}\ket{\psi^{+}}_{CD}\\-\ket{\phi^{+}}_{AC}\ket{\phi^{-}}_{BD}\\-\ket{\phi^{-}}_{AC}\ket{\phi^{+}}_{BD}\big]\\
        \ket{\psi^{-}}_{AC}\otimes\ket{\psi^{+}}_{BD}=\frac{1}{2}\big[\ket{\psi^{+}}_{AB}\ket{\psi^{-}}_{CD}\\+\ket{\psi^{-}}_{AB}\ket{\psi^{+}}_{CD}\\+\ket{\phi^{+}}_{AC}\ket{\phi^{-}}_{BD}\\+\ket{\phi^{-}}_{AC}\ket{\phi^{+}}_{BD}\big]\\
        \ket{\psi^{-}}_{AC}\otimes\ket{\psi^{-}}_{BD}=\frac{1}{2}\big[\ket{\psi^{+}}_{AB}\ket{\psi^{+}}_{CD}\\+\ket{\psi^{-}}_{AB}\ket{\psi^{-}}_{CD}\\-ket{\phi^{+}}_{AC}\ket{\phi^{+}}_{BD}\\-\ket{\phi^{-}}_{AC}\ket{\phi^{-}}_{BD}\big]
    \end{split}
\end{equation}

From Table~\ref{table:2}, we can notice that, after Bell basis measurement is performed on Charlie's qubits, due to entanglement swapping A and B in positions 1, 3, and 4 are entangled. The single qubits in position 2 sent to Charlie are entangled to form Bell state. We can easily see that-

\begin{equation}\label{qsdc_eq4}
        \ket{+-}=\frac{1}{\sqrt{2}}\big(\ket{\psi^{-}}-\ket{\phi^{-}}\big)
\end{equation}

The Bell basis measurement causes Charlie's qubits in position 2 to get entangled into any one of the Bell states given in Eq. (\ref{qsdc_eq4}). The eavesdropper's interception can change the single qubits' states and so a security check is now performed on the Bell state in position 2.

After the security check, the error rate is estimated and if it lies within an acceptable error rate, the communication between Alice and Bob can continue, otherwise, it is terminated. Now, how do we make a security check using quantum circuits to tell whether any eavesdropper has attempted to intercept or not? We developed a technique using quantum circuits through which it can be easily realized. We discuss this in the following subsection.

Bell basis measurement in position 5 and 6 causes the qubits sent to Charlie to become entangled due to entanglement swapping, leaving behind single qubits in A and B which can be understood from the following-

\begin{equation}\
    \begin{split}
        \ket{0}_{C}\otimes\ket{\psi^{-}}_{BD}=\frac{1}{2}\big[\ket{\psi^{+}}_{CD}\ket{0}_{B}\\+\ket{\psi^{-}}_{CD}\ket{0}_{B}\\-\ket{\psi^{+}}_{CD}\ket{1}_{B}\\+\ket{\psi^{-}}_{CD}\ket{1}_{B}\big]\\
        \ket{\psi^{+}}_{AC}\otimes\ket{1}_{D}=\frac{1}{2}\big[\ket{0}_{A}\ket{\phi^{+}}_{CD}\\+\ket{0}_{A}\ket{\phi^{-}}_{CD}\\+\ket{1}_{A}\ket{\psi^{+}}_{CD}\\-\ket{1}_{A}\ket{\phi^{-}}_{CD}\big]
    \end{split}
\end{equation}

For simplicity of the protocol, positions 5 and 6, where single qubits are left behind in A and B after Bell basis measurement, are discarded. 

\subsection{Circuit implementation for Security Check}
For security check, we make use of swapping circuit that we have developed \cite{qsdc_Gupta2020} for calculation of inner product of two qubit states as shown symbolically in the figure below:

\begin{figure}[H]
\centering
\includegraphics[scale=0.3]{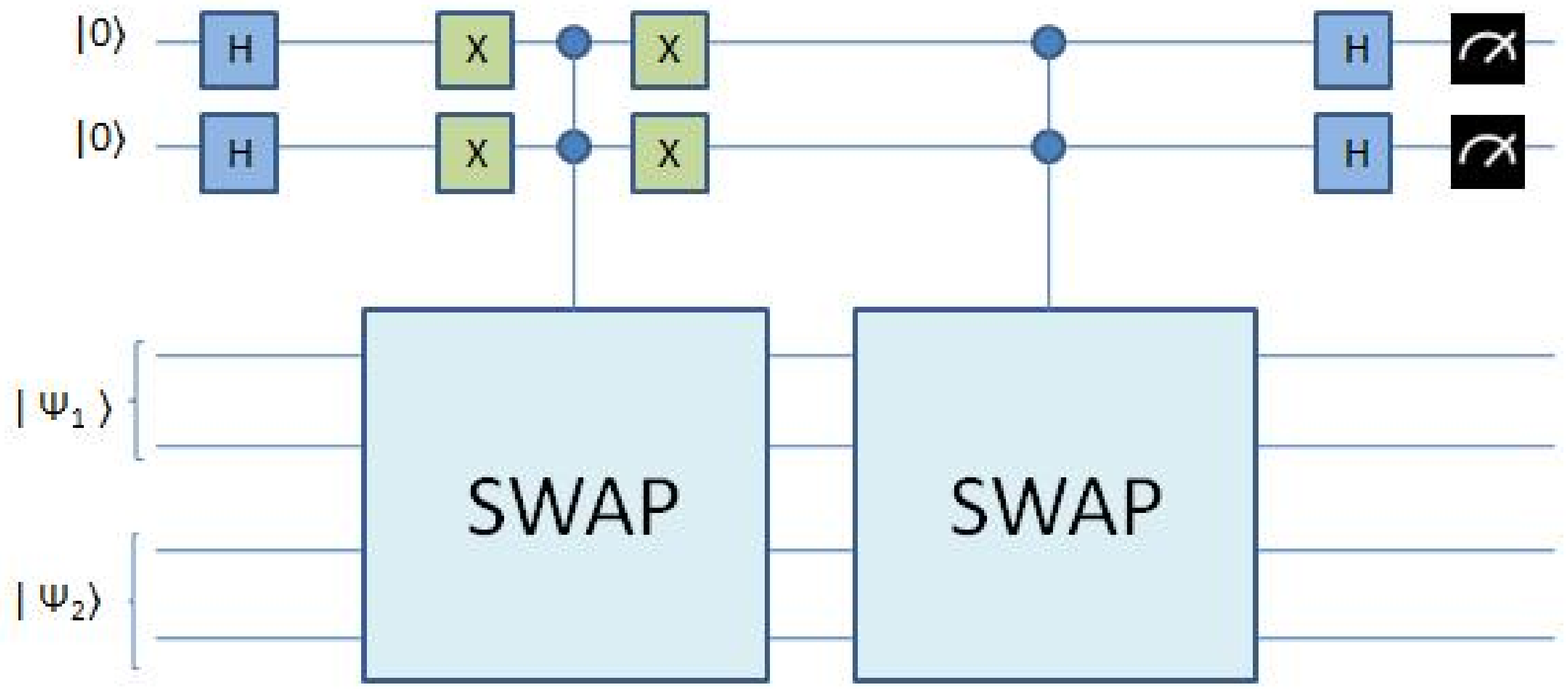}
\caption{Quantum swapping circuit for determining the inner product of two qubit states}
\label{qsdc_Fig1}
\end{figure}

If the projection measurement is performed on the ancilla qubits, the probability that it is in the state $\Ket{00}$ is $\frac{1}{2}(1+|\braket{\psi_1|\psi_2}|^{2})$.

Therefore, the inner product of $\Ket{\psi_1}$ and $\Ket{\psi_2}$ can be calculated as $\braket{\psi_1|\psi_2}=\sqrt{2P(\Ket{00})-1}$, where $P(\Ket{00})$ is the probability that the ancilla qubits are in state $\Ket{00}$.

We calculate the inner product of the initial qubits sent to Charlie $\ket{+}$ and $\ket{-}$ with the entangled state in Charlie's possession. If the inner product is not within acceptable error (this implies an interception attempt by an eavesdropper), the communication between Alice and Bob is terminated. Otherwise, the communication process is allowed to proceed. The circuit implemented for calculating the inner product is shown in the figure below-

\begin{figure}[H]
\centering
\includegraphics[scale=0.3]{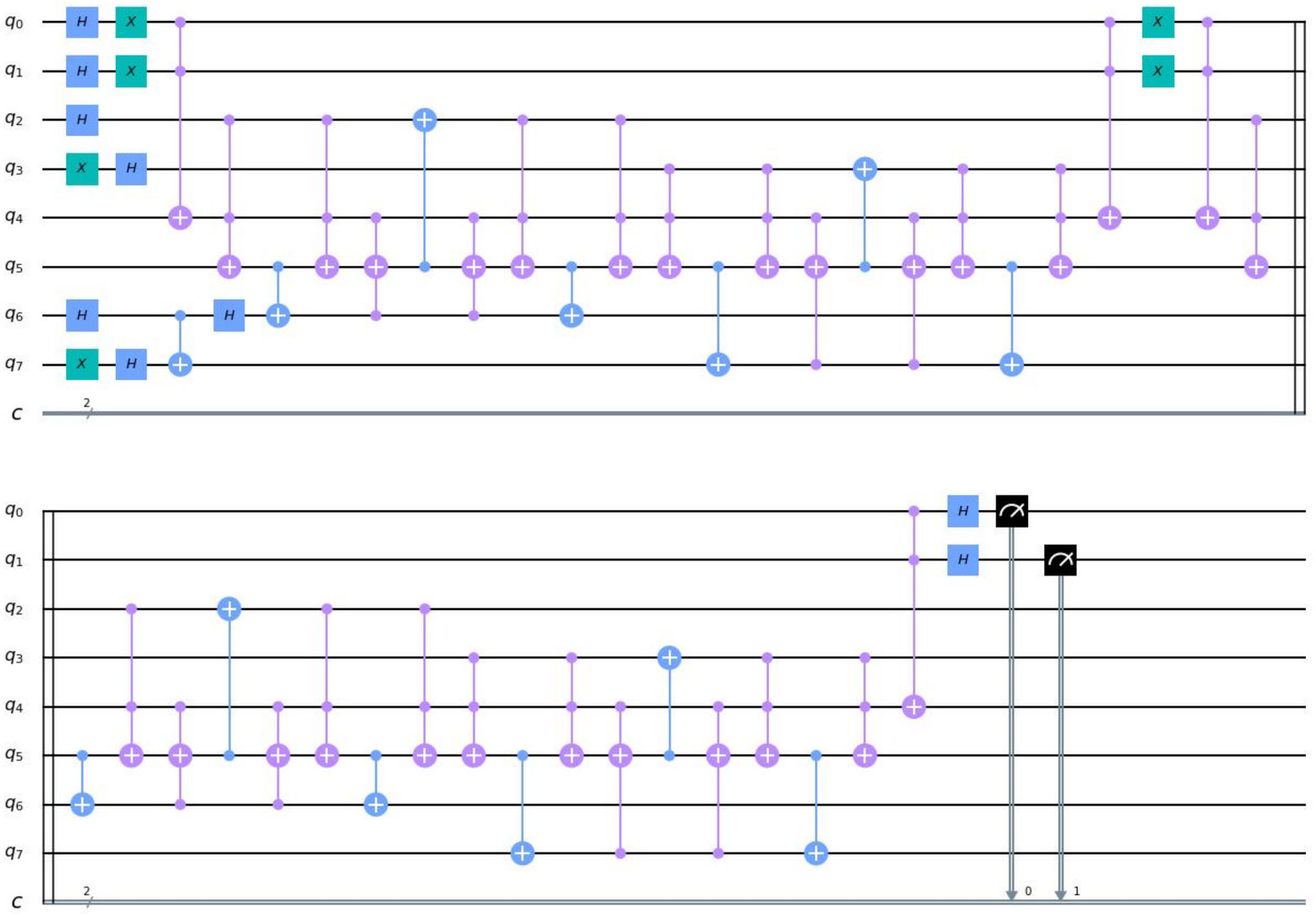}
\caption{Circuit implemented for security check without eavesdropper interception}
\label{qsdc_Fig2}
\end{figure}

The projection measurement on the ancilla qubits gives the following result-

\begin{figure}[H]
\centering
\includegraphics[scale=0.3]{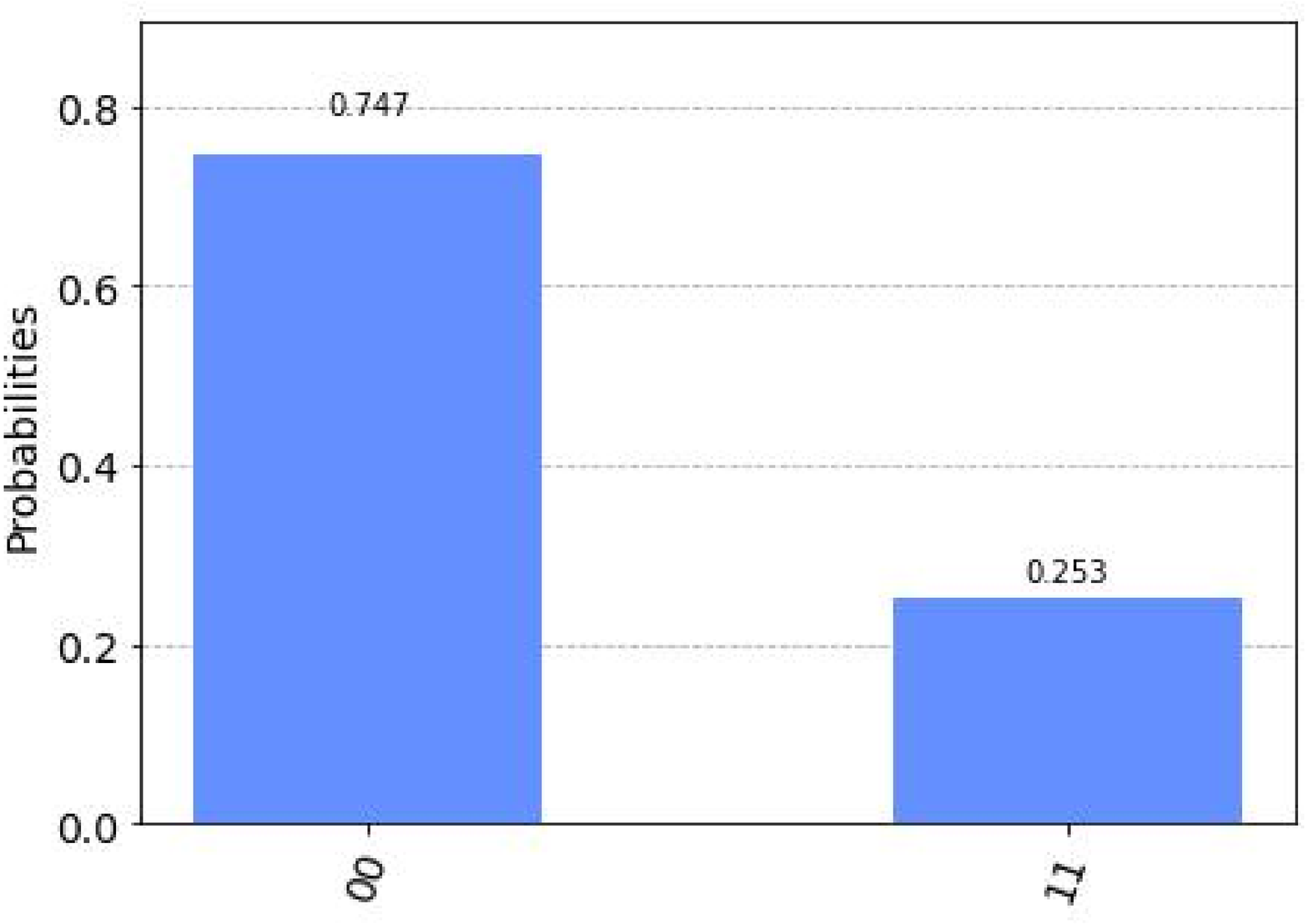}
\caption{Histogram plot of the result for Run 1}
\label{qsdc_Fig3}
\end{figure}
The value of the inner product is given by-
\begin{equation*}
    \sqrt{2(P\ket{00})-1}=\sqrt{2\times0.747-1}=0.70285
\end{equation*}
We make two more runs for this circuit in order to set an acceptable error range-
\begin{figure}[H]
\centering
\includegraphics[scale=0.3]{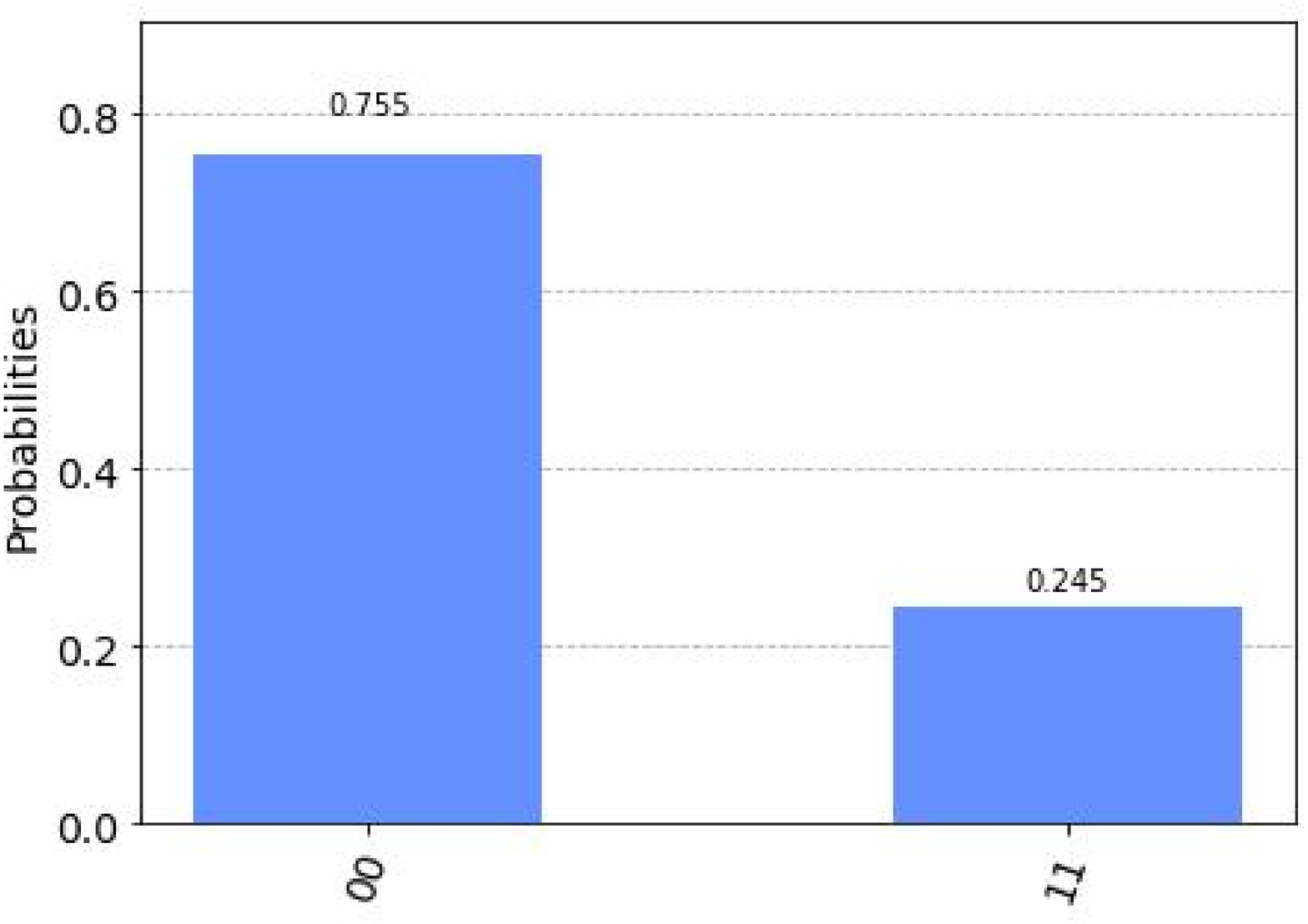}
\caption{Histogram plot of the result for Run 2}
\label{qsdc_Fig4}
\end{figure}
The value of the inner product is given by-
\begin{equation*}
    \sqrt{2(P\ket{00})-1}=\sqrt{2\times0.755-1}=0.71414
\end{equation*}
\begin{figure}[H]
\centering
\includegraphics[scale=0.3]{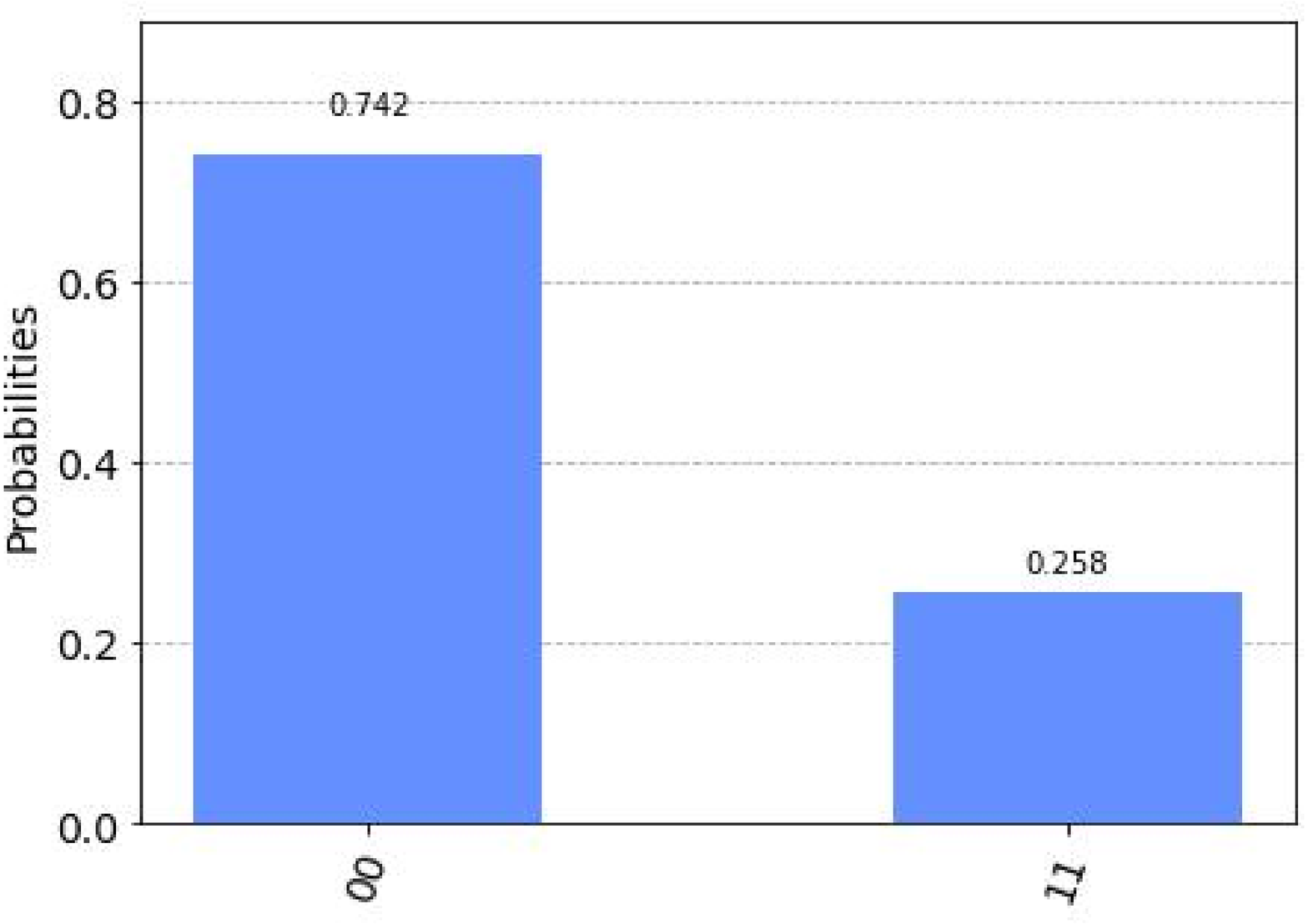}
\caption{Histogram plot of the result for Run 3}
\label{qsdc_Fig5}
\end{figure}
The value of the inner product is given by-
\begin{equation*}
    \sqrt{2(P\ket{00})-1}=\sqrt{2\times0.742-1}=0.69570
\end{equation*}

If the eavesdropper intercepts, this will lead to the 1-qubit states that were randomly inserted into C and D in position 2 to collapse into some other state and hence by calculating the inner product, we can easily detect any attempt of eavesdropping. Let us suppose that $\ket{+}$ in C changes to $\ket{0}$ due to interception by the eavesdropper before Bell basis measurement is made. The circuit implemented to demonstrate the above is shown in the following figure-

\begin{figure}[H]
\centering
\includegraphics[scale=0.3]{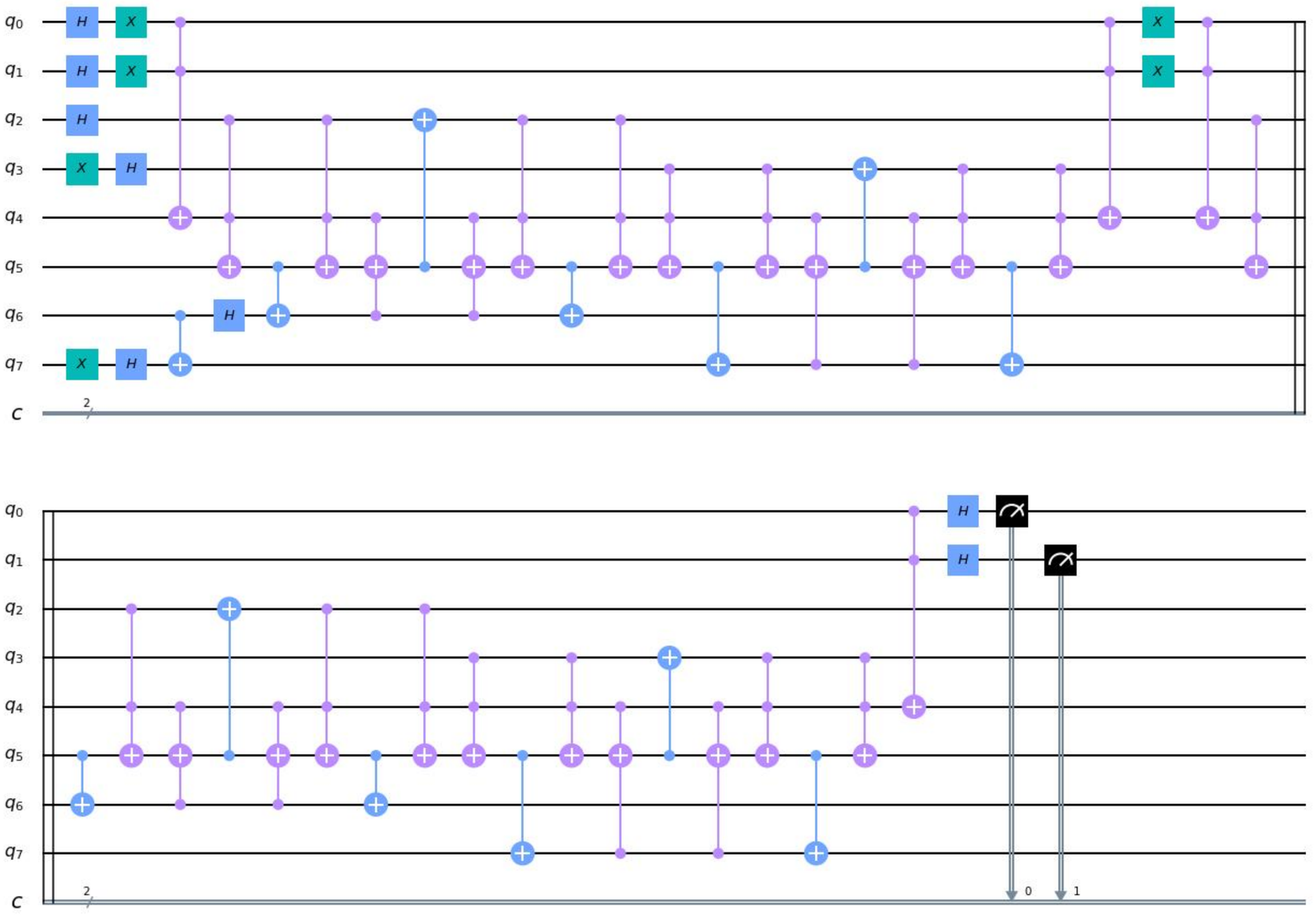}
\caption{Circuit implemented for security check with eavesdropper interception}
\label{qsdc_Fig6}
\end{figure}
The projection measurement on the ancilla qubits gives us the following result-
\begin{figure}[H]
\centering
\includegraphics[scale=0.3]{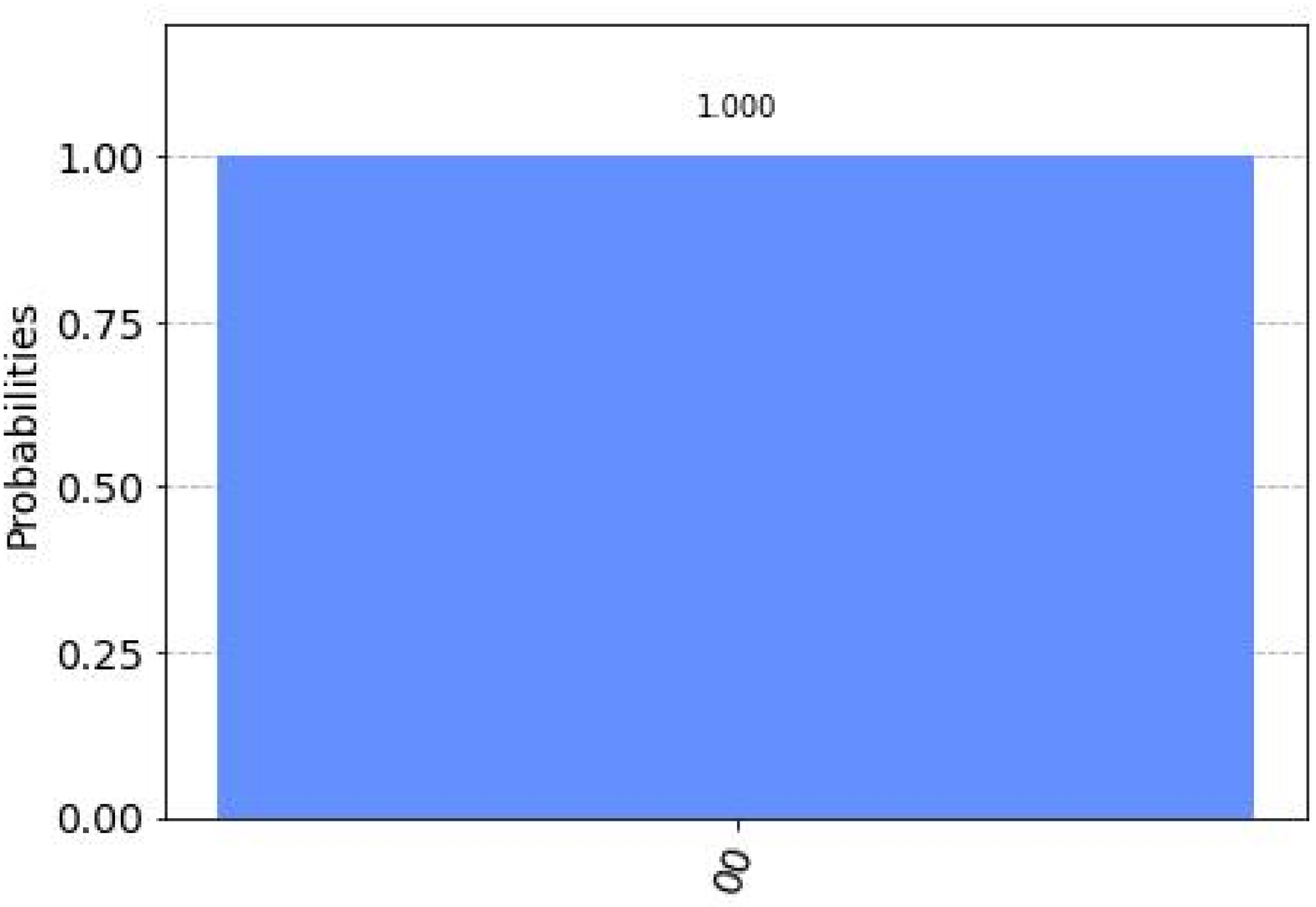}
\caption{Result of the measurement on ancilla qubits}
\label{qsdc_Fig7}
\end{figure}
The value of the inner product is given by-
\begin{equation*}
    \sqrt{2(P\ket{00})-1}=\sqrt{2\times1-1}=1.00000
\end{equation*}
Again, due to eavesdropping suppose that $\ket{-}$ in D changes to $\ket{0}$ before the Bell basis measurement is made. The circuit implemented to demonstrate the above is shown in the following figure-

\begin{figure}[H]
\centering
\includegraphics[scale=0.3]{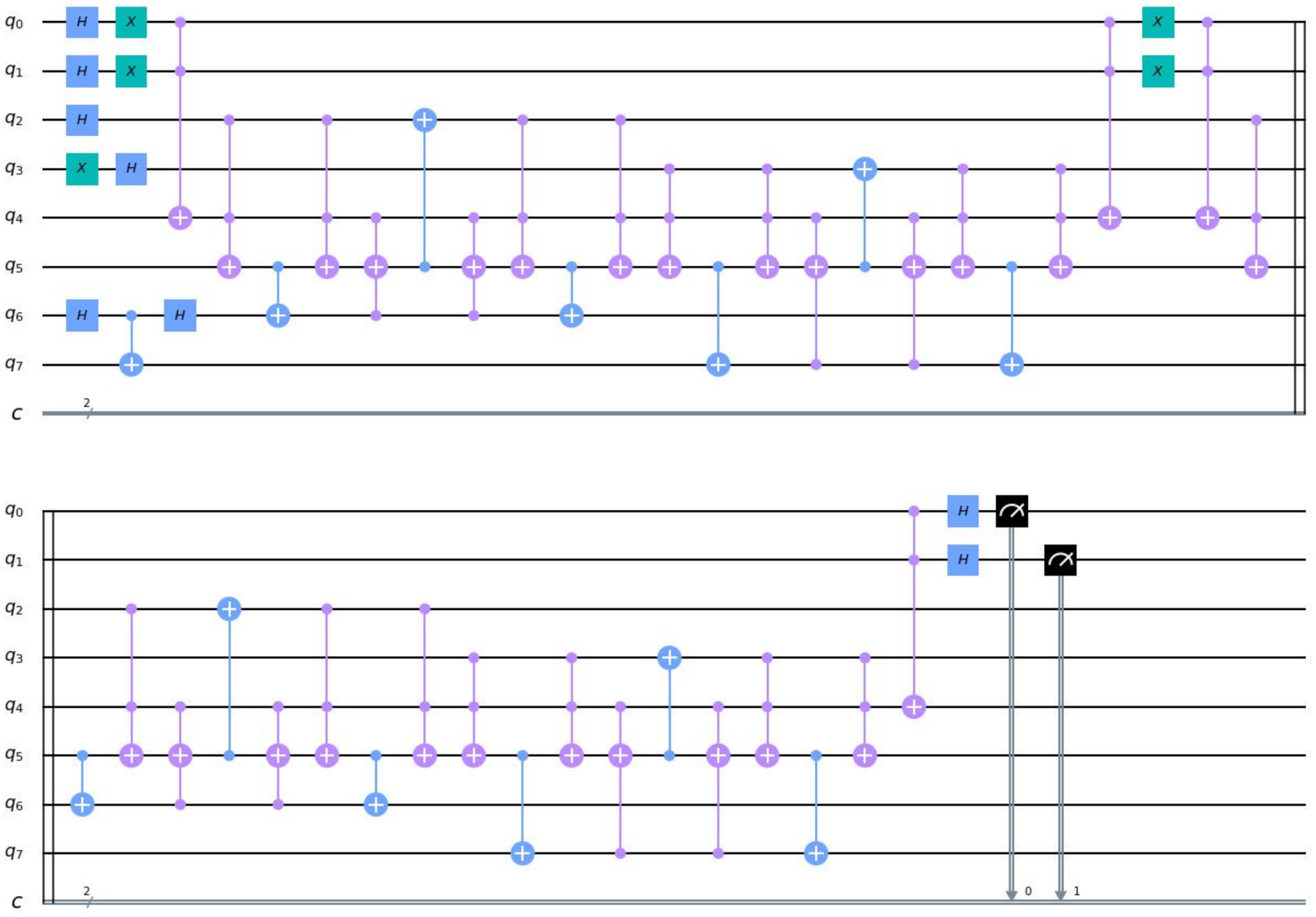}
\caption{Circuit implemented for security check with eavesdropper interception}
\label{qsdc_Fig8}
\end{figure}
The projection measurement on the ancilla qubits gives the following result-
\begin{figure}[H]
\centering
\includegraphics[scale=0.3]{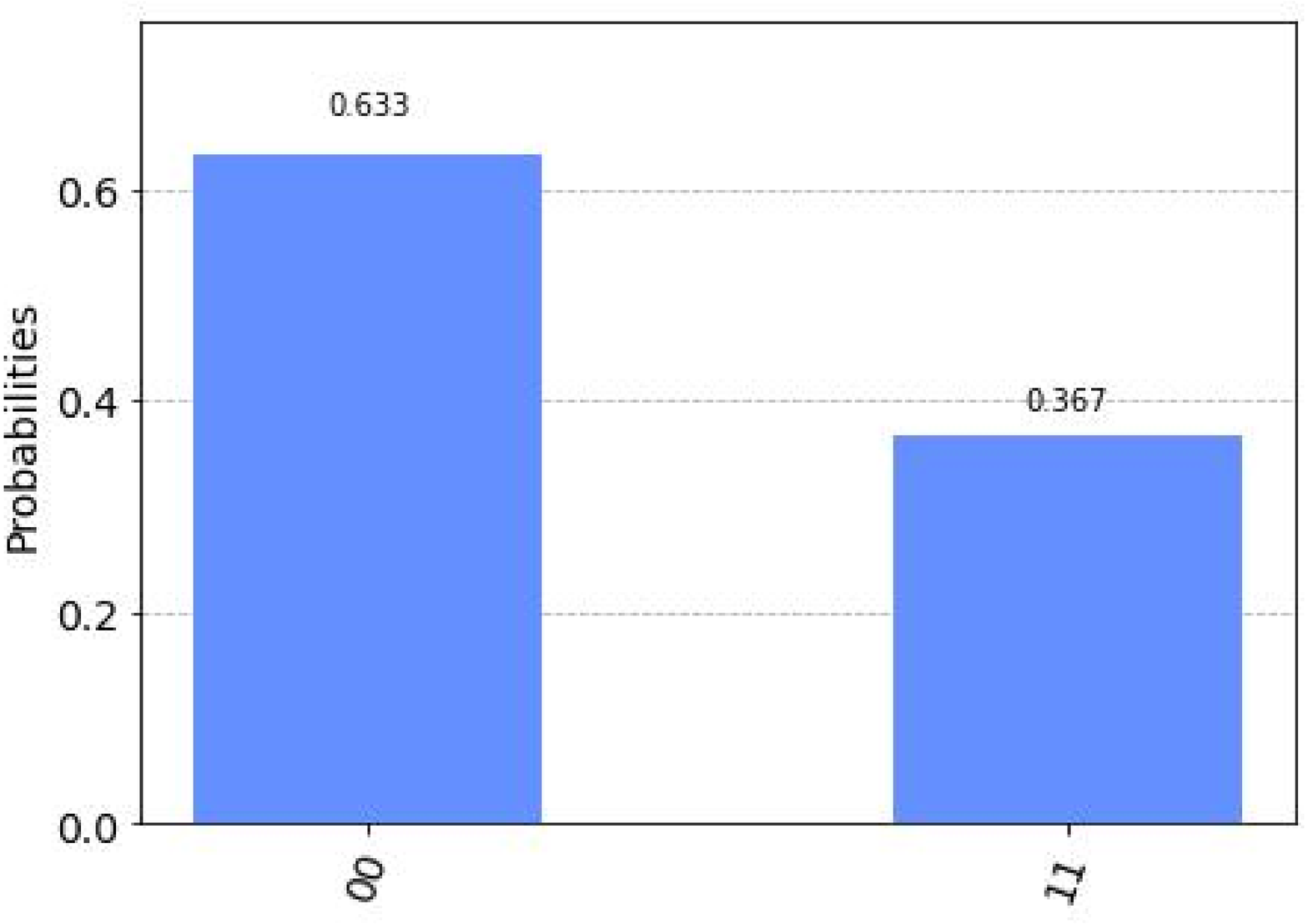}
\caption{Result of the measurement on ancilla qubits}
\label{qsdc_Fig9}
\end{figure}
The value of the inner product is given by-
\begin{equation*}
    \sqrt{2(P\ket{00})-1}=\sqrt{2\times0.633-1}=0.51575
\end{equation*}
Now, we make a table listing the error rates and deciding whether to terminate or proceed with the communication process.

\setlength{\arrayrulewidth}{0.5mm}
\setlength{\tabcolsep}{15pt}
\renewcommand{\arraystretch}{3.5}
\begin{table}[H]
\caption{Security Check}
\begin{tabular}{|c|c|c|c|}
\hline
S.No. &Value   &Error rate  & Comm. status \\ \hline
1     &0.70285 &-           &Proceed \\
2     &0.71414 &$0.016\%$   &Proceed \\
3     &0.69570 &$-0.010\%$  &Proceed \\
4     &1.00000 &$0.423\%$   &Terminate \\
5     &0.51575 &$-0.266\%$  &Terminate \\ \hline
\end{tabular}
\label{table:3}
\end{table}

We observe that the inner product values whose error rates are very low below a standard cutoff value, the communication proceeds, otherwise it is immediately terminated.

\subsection{Circuit implementation for Entanglement Swapping}
We have already discussed how Bell basis measurements on Charlie's qubits leads to entangled pairs for Alice and Bob due to entanglement swapping. These pairs in A and B form an ordered sequence $M_A$ and $M_B$ respectively. Now in order to send a message we apply a Z gate on Alice's qubit whose initial states are $\ket{\psi^{+}}$. From Eq. (\ref{qsdc_eq3}), we can see that this is equivalent to preparing all the initial states of Alice in the $\ket{\psi^{-}}$ state. So, $M_A$ only contains qubits whose initial states are $\ket{\psi^{-}}$ while $M_B$ contains qubits from both $\ket{\psi^{+}}$ and $\ket{\psi^{-}}$ states. The circuits implemented to realize the same is given in the figures below-

\begin{figure}[H]
\centering
\includegraphics[scale=0.3]{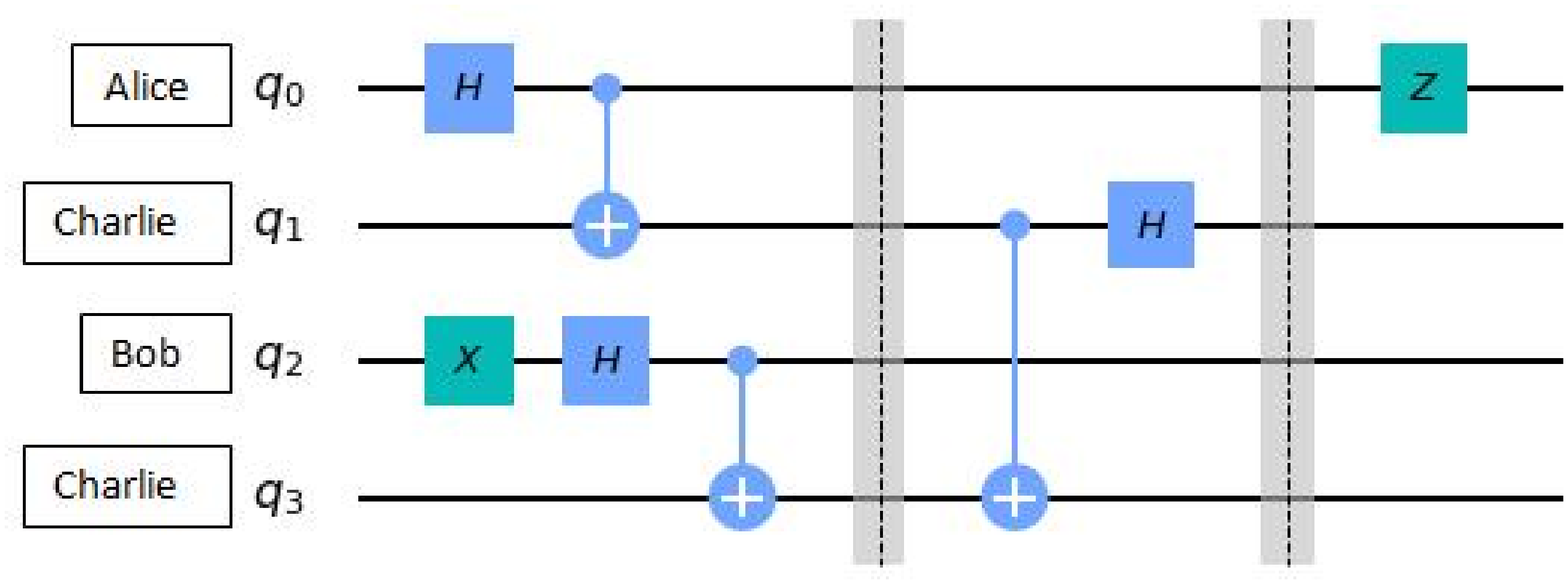}
\caption{Circuit implementation for entangled state 1}
\label{qsdc_Fig10}
\end{figure}
\begin{figure}[H]
\centering
\includegraphics[scale=0.3]{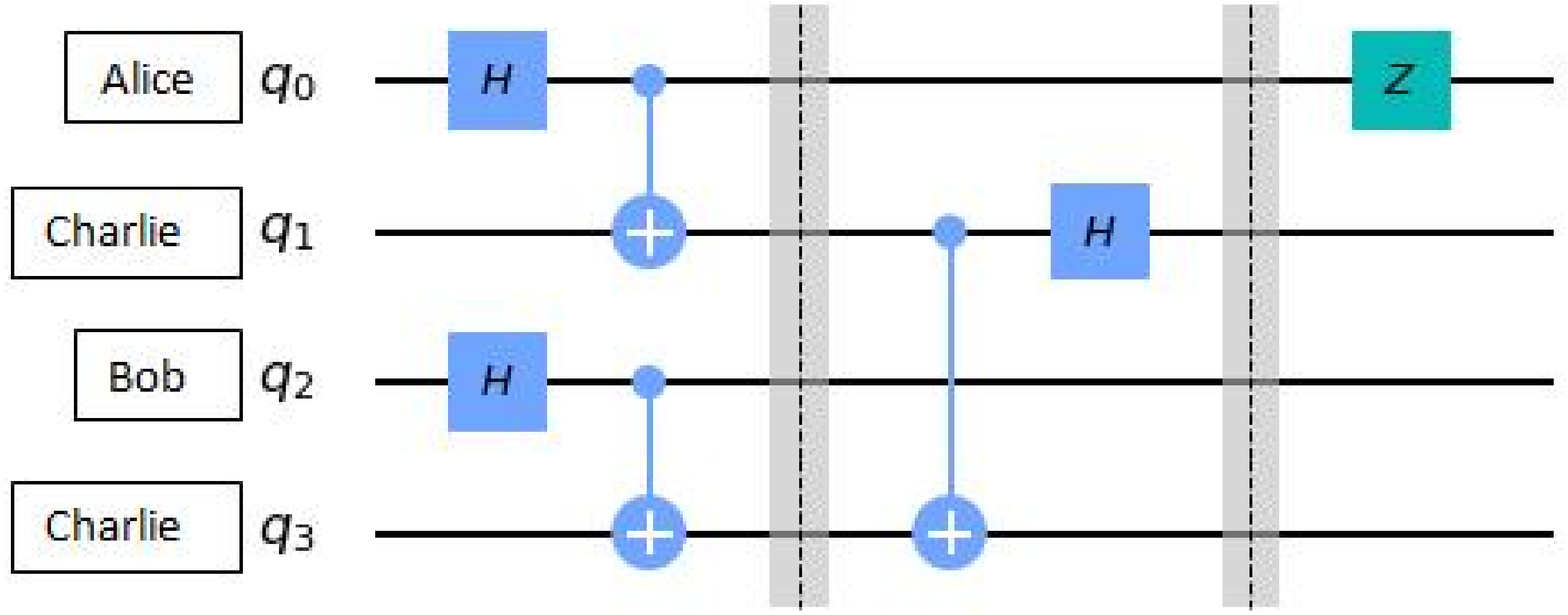}
\caption{Circuit implementation for entangled state 2}
\label{qsdc_Fig11}
\end{figure}
\begin{figure}[H]
\centering
\includegraphics[scale=0.3]{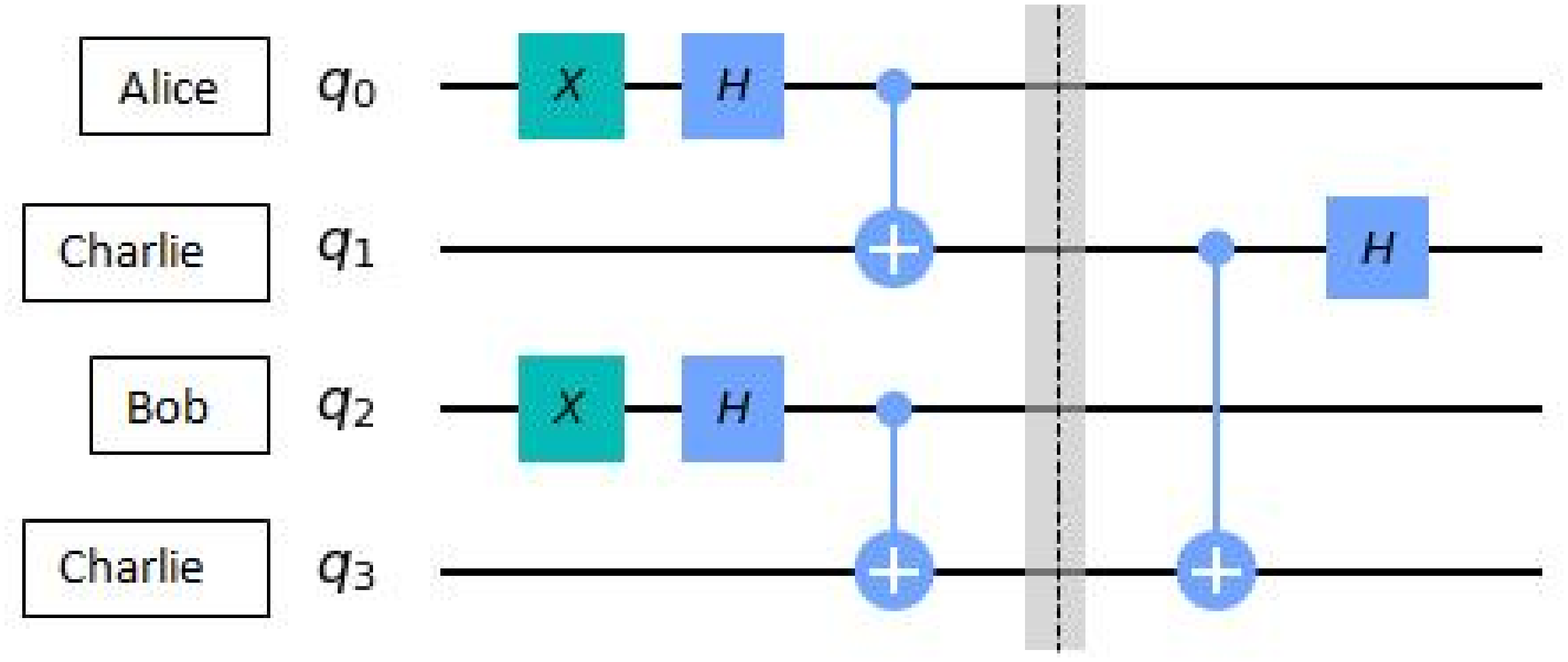}
\caption{Circuit implementation for entangled state 3}
\label{qsdc_Fig12}
\end{figure}

\subsection{Circuit implementation for Superdense coding}
We assume that after Bell basis measurement, due to entanglement swapping the entangled state between Alice and Bob is the Bell state $\ket{\psi^{+}}_{AB}$. We perform superdense coding operations on Alice's qubits in $M_A$ and randomly apply Z gates on Bob's qubits in $M_B$ to prevent the eavesdropper from performing the intercept and resend attack. $M_A$ and $M_B$ are now sent to Charlie, the third party, who now performs Bell basis measurement and publishes the results. Now, we implement the circuits for superdense coding operations with and without applying Z gate on Bob's qubits.

\begin{figure}[H]
\centering
\includegraphics[scale=0.3]{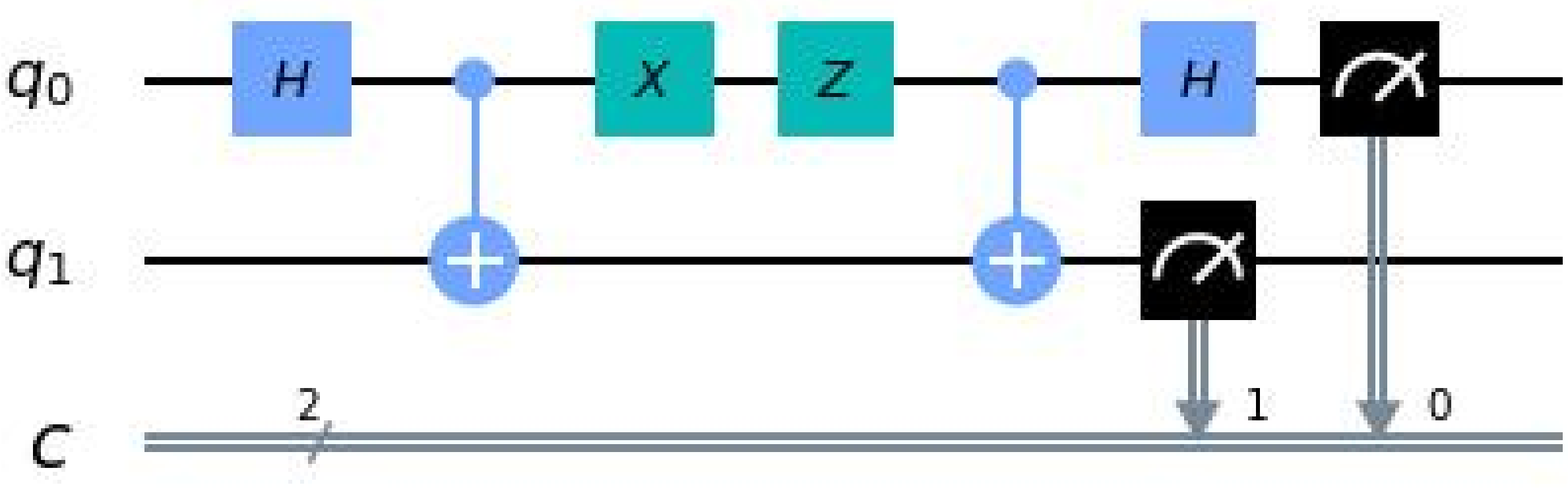}
\caption{Superdense coding circuit without applying Z gate on Bob's qubit}
\label{qsdc_Fig13}
\end{figure}
On plotting the histogram for the probability of the qubits received by Bob, we get-
\begin{figure}[H]
\centering
\includegraphics[scale=0.3]{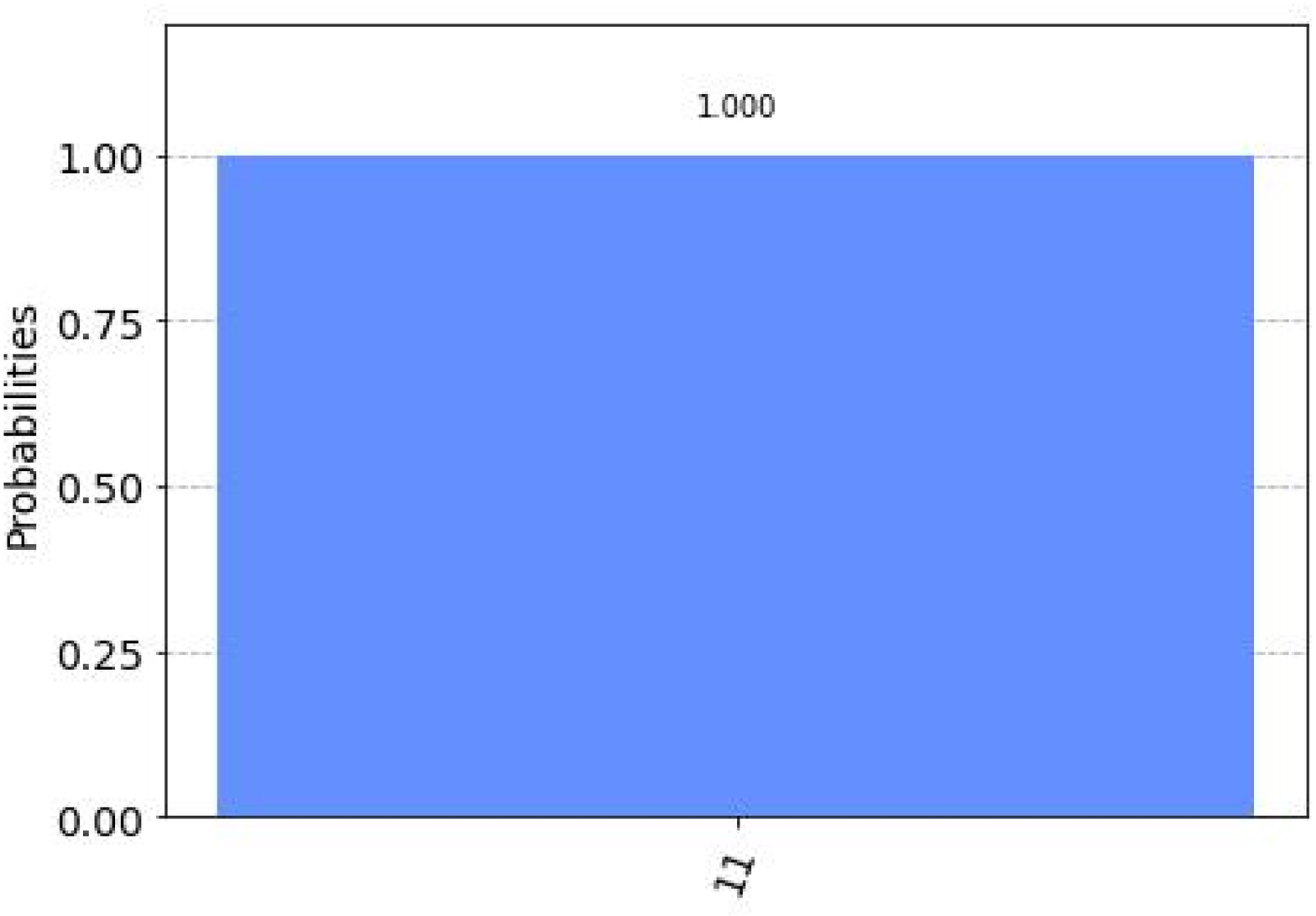}
\caption{Histogram plot of the message received by Bob}
\label{qsdc_Fig14}
\end{figure}
Again, now after applying Z gate on Bob's qubit we can see how the message received by Bob changes.
\begin{figure}[H]
\centering
\includegraphics[scale=0.3]{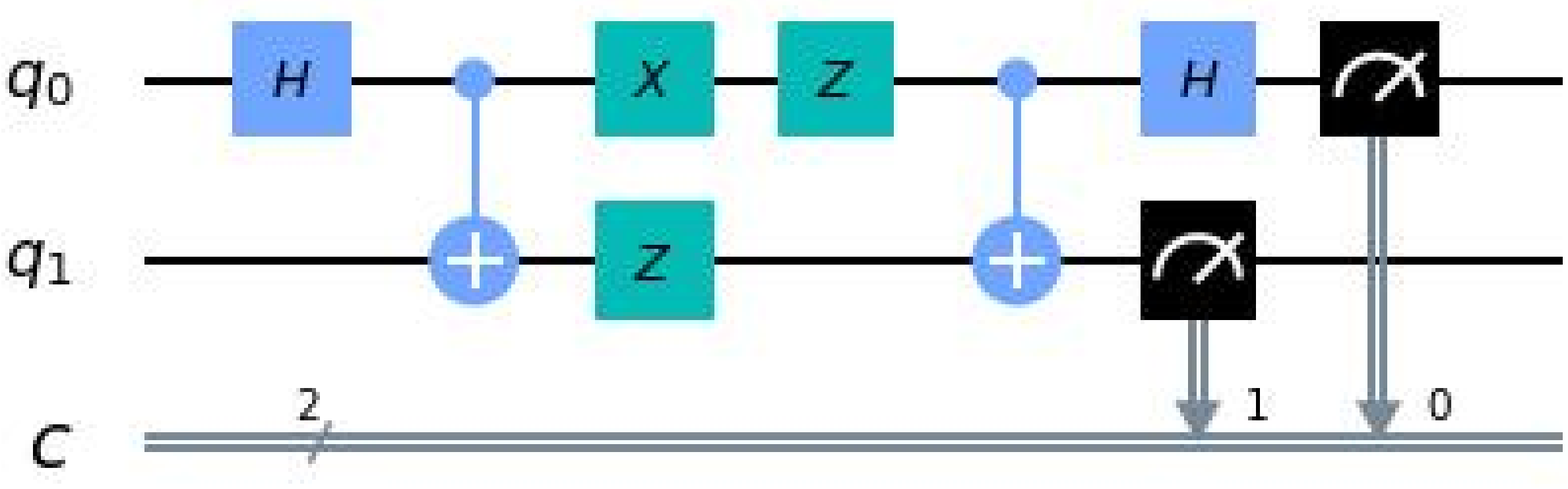}
\caption{Superdense coding circuit after applying Z gate on Bob's qubit}
\label{qsdc_Fig15}
\end{figure}
Again, after plotting the histogram, we find the message received by Bob.
\begin{figure}[H]
\centering
\includegraphics[scale=0.3]{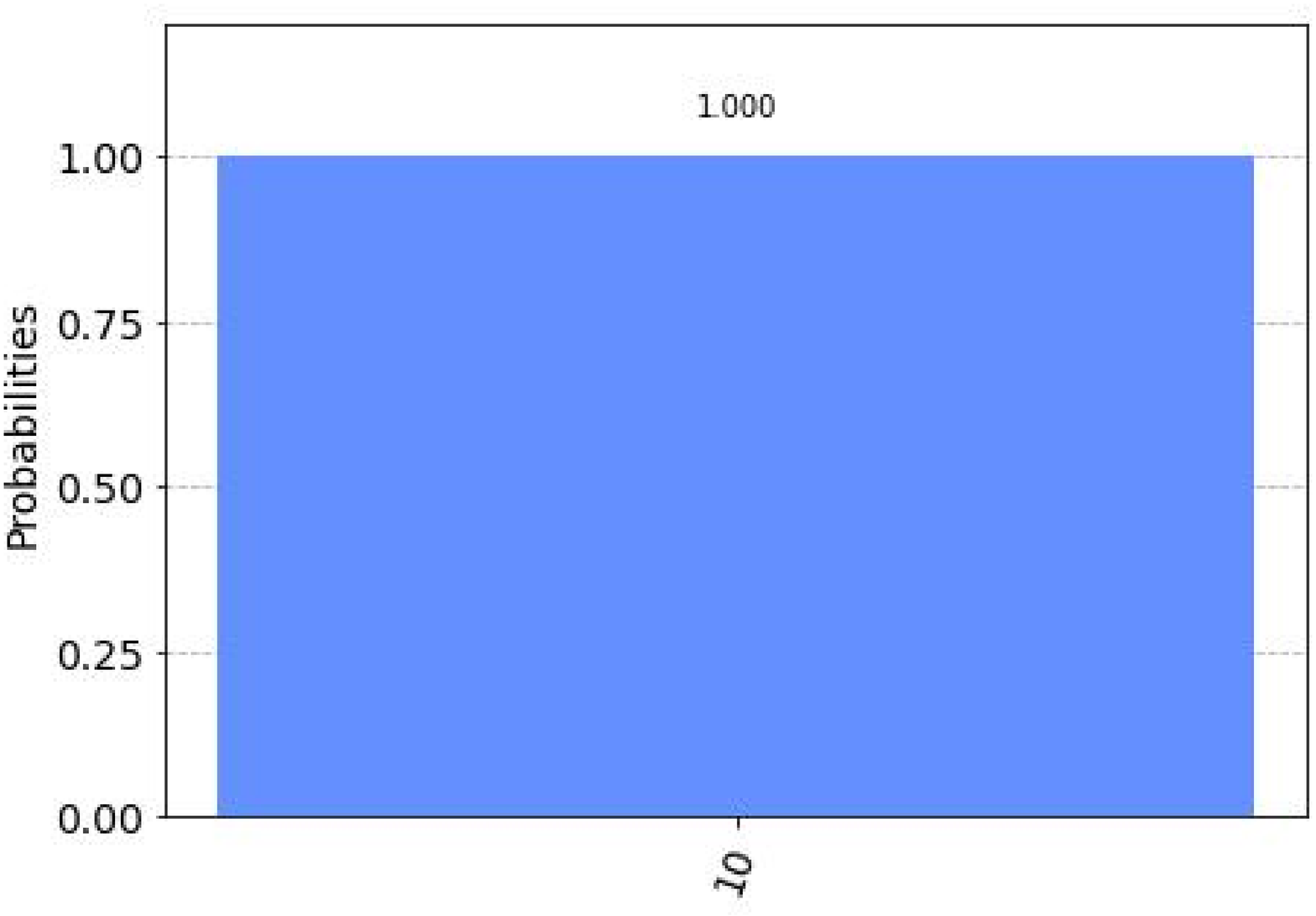}
\caption{Histogram plot of the message received by Bob}
\label{qsdc_Fig16}
\end{figure}

Thus we notice that $\ket{11}$ becomes $\ket{01}$ after Z gate is applied on Bob's qubit. Similarly, we can check it for the other three superdense coding operations. Since it is known only to Bob where Bob has applied Z gates, Charlie cannot know the secret message sent by Alice to Bob. After the results are published by Charlie, Bob compares it with Alice to decode the message.

\section{MDI-QSDC protocol using GHZ states \label{qsdc_Sec3}}
Greenberger-Horne-Zeilinger (GHZ) states are given by-

\begin{equation}
    \begin{split}
        \ket{\psi_{000}}=\frac{1}{\sqrt{2}}\big[\ket{000}+\ket{111}\big]\\
        \ket{\psi_{001}}=\frac{1}{\sqrt{2}}\big[\ket{000}-\ket{111}\big]\\
        \ket{\psi_{010}}=\frac{1}{\sqrt{2}}\big[\ket{100}+\ket{011}\big]\\
        \ket{\psi_{011}}=\frac{1}{\sqrt{2}}\big[\ket{100}-\ket{011}\big]\\
        \ket{\psi_{100}}=\frac{1}{\sqrt{2}}\big[\ket{010}+\ket{101}\big]\\
        \ket{\psi_{101}}=\frac{1}{\sqrt{2}}\big[\ket{010}-\ket{101}\big]\\
        \ket{\psi_{110}}=\frac{1}{\sqrt{2}}\big[\ket{110}+\ket{001}\big]\\
        \ket{\psi_{111}}=\frac{1}{\sqrt{2}}\big[\ket{110}-\ket{001}\big]
    \end{split}
\end{equation}

The protocol is basically similar to that of Bell state except that two qubits are kept in Alice's possession and one is sent to Charlie while one qubit is kept in Bob's possession and two are sent to Charlie for Charlie to perform GHZ basis measurement. We prepare Alice's and Bob's state in the $\ket{\psi_{000}}$ and $\ket{\psi_{001}}$ states and then single qubit states in computational and Hadamard basis are inserted in random positions for Alice and two qubit states for Bob which are sent to Charlie, the third party.

\setlength{\arrayrulewidth}{0.5mm}
\setlength{\tabcolsep}{28pt}
\renewcommand{\arraystretch}{2.0}
\begin{table}[H]
\caption{Random insertion of qubits}
\begin{tabular}{|c|c|c|}
\hline
Position  &Alice               & Bob \\ \hline
1         &$\ket{\psi_{000}}$ &$\ket{\psi_{000}}$ \\
2         &$\ket{+}$          &$\ket{+-}$\\
3         &$\ket{\psi_{000}}$ &$\ket{\psi_{000}}$ \\
4         &$\ket{\psi_{001}}$ &$\ket{\psi_{001}}$ \\
5         &$\ket{0}$          &$\ket{\psi_{000}}$ \\ 
6         &$\ket{\psi_{000}}$ &$\ket{00}$\\ \hline
\end{tabular}
\label{table:4}
\end{table}

We assign two of the qubits for Alice and one qubit for Bob of the GHZ states at different positions denoted by $A_1$ and $A_2$ for Alice and B for Bob. The other qubit from Alice's GHZ state along with the randomly inserted single qubit states are sent to Charlie denoted by $C_1$ while in Bob's side the other two qubits of the GHZ state along with the randomly inserted two qubit states are sent to Charlie denoted by $C_2$ and $C_3$. GHZ basis measurements are now performed on the qubits sent to Charlie. Charlie announces the results of the GHZ basis measurements. Due to entanglement swapping, the qubits in $A_1$, $A_2$ and B become entangled-

\begin{equation}\label{qsdc_eq7}
    \begin{split}
        \ket{\psi_{000}}_{A_{1}A_{2}C_{1}}\otimes\ket{\psi_{000}}_{BC_{2}C_{3}}\\=\frac{1}{2}\big[\ket{\psi_{000}}_{A_{1}A_{2}B}\ket{\psi_{000}}_{C_{1}C_{2}C_{3}}\\+\ket{\psi_{000}}_{A_{1}A_{2}B}\ket{\psi_{001}}_{C_{1}C_{2}C_{3}}\\+\ket{\psi_{110}}_{A_{1}A_{2}B}\ket{\psi_{010}}_{C_{1}C_{2}C_{3}}\\+\ket{\psi_{110}}_{A_{1}A_{2}B}\ket{\psi_{011}}_{C_{1}C_{2}C_{3}}\big]\\\\
        \ket{\psi_{001}}_{A_{1}A_{2}C_{1}}\otimes\ket{\psi_{001}}_{BC_{2}C_{3}}\\=\frac{1}{2}\big[\ket{\psi_{000}}_{A_{1}A_{2}B}\ket{\psi_{000}}_{C_{1}C_{2}C_{3}}\\+\ket{\psi_{000}}_{A_{1}A_{2}B}\ket{\psi_{001}}_{C_{1}C_{2}C_{3}}\\-\ket{\psi_{110}}_{A_{1}A_{2}B}\ket{\psi_{010}}_{C_{1}C_{2}C_{3}}\\-\ket{\psi_{110}}_{A_{1}A_{2}B}\ket{\psi_{011}}_{C_{1}C_{2}C_{3}}\big]
    \end{split}
\end{equation}
From Table \ref{table:4}, we can notice that, after GHZ basis measurement is performed on Charlie's qubits, due to entanglement swapping $A_1$, $A_2$ and B in position 1, 3 and 4 are entangled. The one and two qubit states in position 2 sent to Charlie are entangled to form GHZ state. We can see that-
\begin{equation}\label{qsdc_eq8}
    \begin{split}
        \ket{+}_{C_{1}}\otimes\ket{+-}_{C_{2}C_{3}}=\frac{1}{2}\big[\ket{\psi_{001}}_{C_{1}C_{2}C_{3}}\\+\ket{\psi_{111}}_{C_{1}C_{2}C_{3}}\\+\ket{\psi_{101}}_{C_{1}C_{2}C_{3}}\\\ket{\psi_{011}}_{C_{1}C_{2}C_{3}}\big]
    \end{split}
\end{equation}

The GHZ basis measurement causes Charlie's qubits in position 2 to get entangled into any of the GHZ states given in Eq. (\ref{qsdc_eq8}). The eavesdropper's interception can change the state of the qubits in position 2 and hence a security check is now performed on lines similar to security check in the previous section. GHZ basis measurement in position 5 and 6 causes the qubits sent to Charlie to become entangled due to entanglement swapping, leaving behind two qubits in $A_1$ and $A_2$ and single qubit in B, which can be understood from the following-

\begin{equation}
   \begin{split}
        \ket{0}_{C_{1}}\otimes\ket{\psi_{000}}_{BC_{2}C_{3}}\\=\frac{1}{2}\big[\ket{0}_{B}\ket{\psi_{000}}_{C_{1}C_{2}C_{3}}\\+\ket{0}_{B}\ket{\psi_{001}}_{C_{1}C_{2}C_{3}}\\+\ket{1}_{B}\ket{\psi_{010}}_{C_{1}C_{2}C_{3}}\\-\ket{1}_{B}\ket{\psi_{011}}_{C_{1}C_{2}C_{3}}\big]\\\\
        \ket{\psi_{000}}_{A_{1}A_{2}C_1}\otimes\ket{00}_{C_{2}C_{3}}\\=\frac{1}{2}\big[\ket{00}_{A_{1}A_{2}}\ket{\psi_{000}}_{C_{1}C_{2}C_{3}}\\+\ket{00}_{A_{1}A_{2}}\ket{\psi_{001}}_{C_{1}C_{2}C_{3}}\\+\ket{11}_{A_{1}A_{2}}\ket{\psi_{010}}_{C_{1}C_{2}C_{3}}\\+\ket{11}_{A_{1}A_{2}}\ket{\psi_{011}}_{C_{1}C_{2}C_{3}}\big]
    \end{split} 
\end{equation}

For simplicity of the protocol, the single qubit left behind in B and the two qubits in $A_1$ and $A_2$, in position 5 and 6 respectively, are discarded.

\subsection{Circuit implementation for Security Check}
For security check yet again, we make use of swapping circuit, now this time for calculation of three-qubit states as shown symbolically in the figure below:

\begin{figure}[H]
\centering
\includegraphics[scale=0.3]{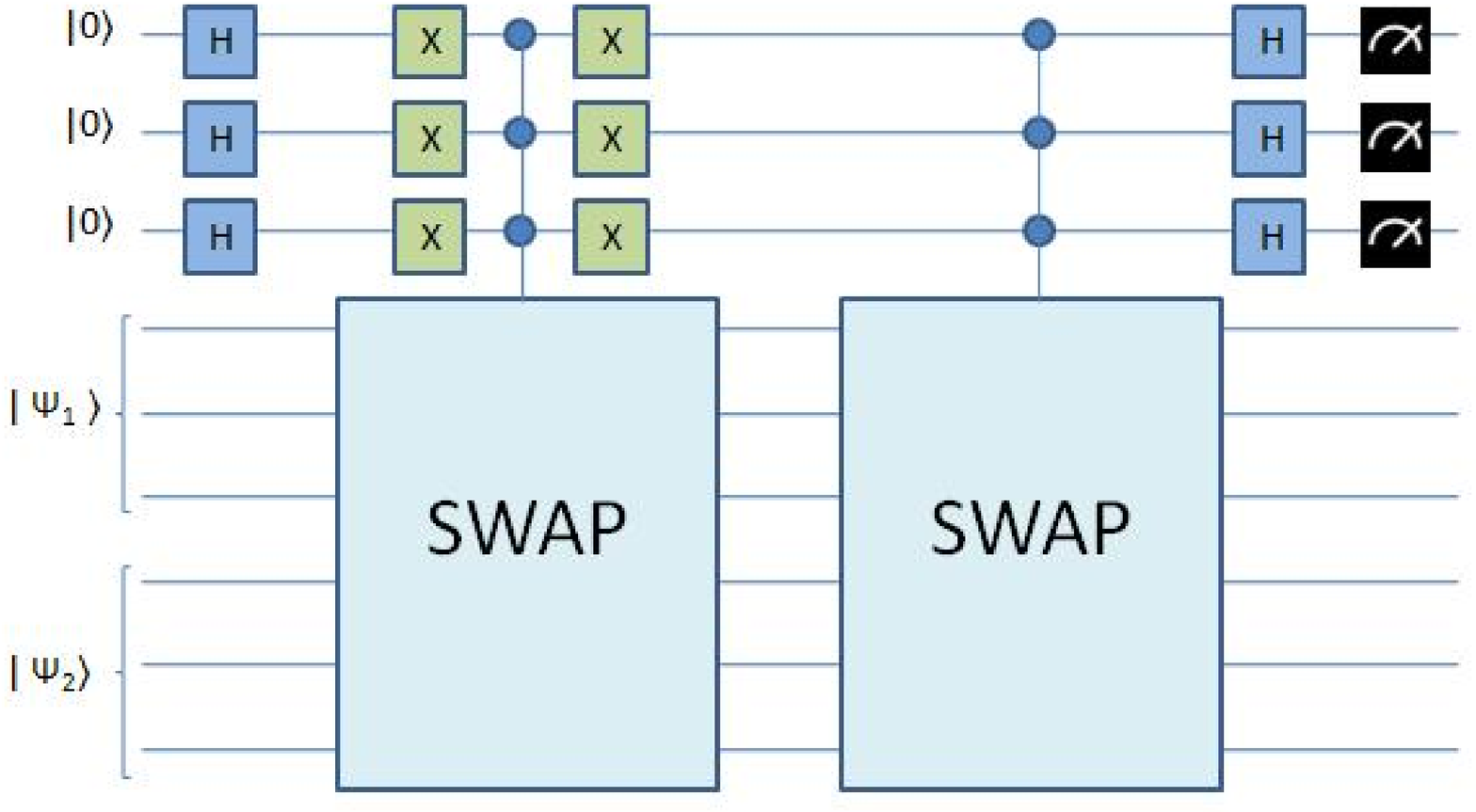}
\caption{Quantum swapping circuit for determining the inner product of three-qubit states}
\label{qsdc_Fig17}
\end{figure}

On similar lines, as in the previous section, we calculate the inner product of the initial qubits sent to Charlie $\ket{+}$ and $\ket{+-}$ with the entangled state in Charlie's possession. If the inner product is not within acceptable error (this implies an interception attempt by an eavesdropper), the communication process between Alice and Bob is terminated. Otherwise, the communication process is allowed to proceed.

\subsection{Circuit implementation for Entanglement Swapping}
We have already discussed how GHZ basis measurement on Charlie's qubits leads to entanglement between Alice's and Bob's qubits. Each party form an ordered sequence $M_A$ and $M_B$ respectively. Now, in order to send a message we apply Z gate on anyone of Alice's two qubits whose initial states are $\ket{\psi_{001}}$. From Eq. (\ref{qsdc_eq7}), we can see that this is equivalent to preparing all the initial states of Alice in the $\ket{\psi_{000}}$ state. So, $M_A$ only contains qubits whose initial states are $\ket{\psi_{000}}$ while $M_B$ contains qubits from both $\ket{\psi_{000}}$ and $\ket{\psi_{001}}$ states. A circuit implemented for the realization of the same is shown in figure below-

\begin{figure}[H]
\centering
\includegraphics[scale=0.3]{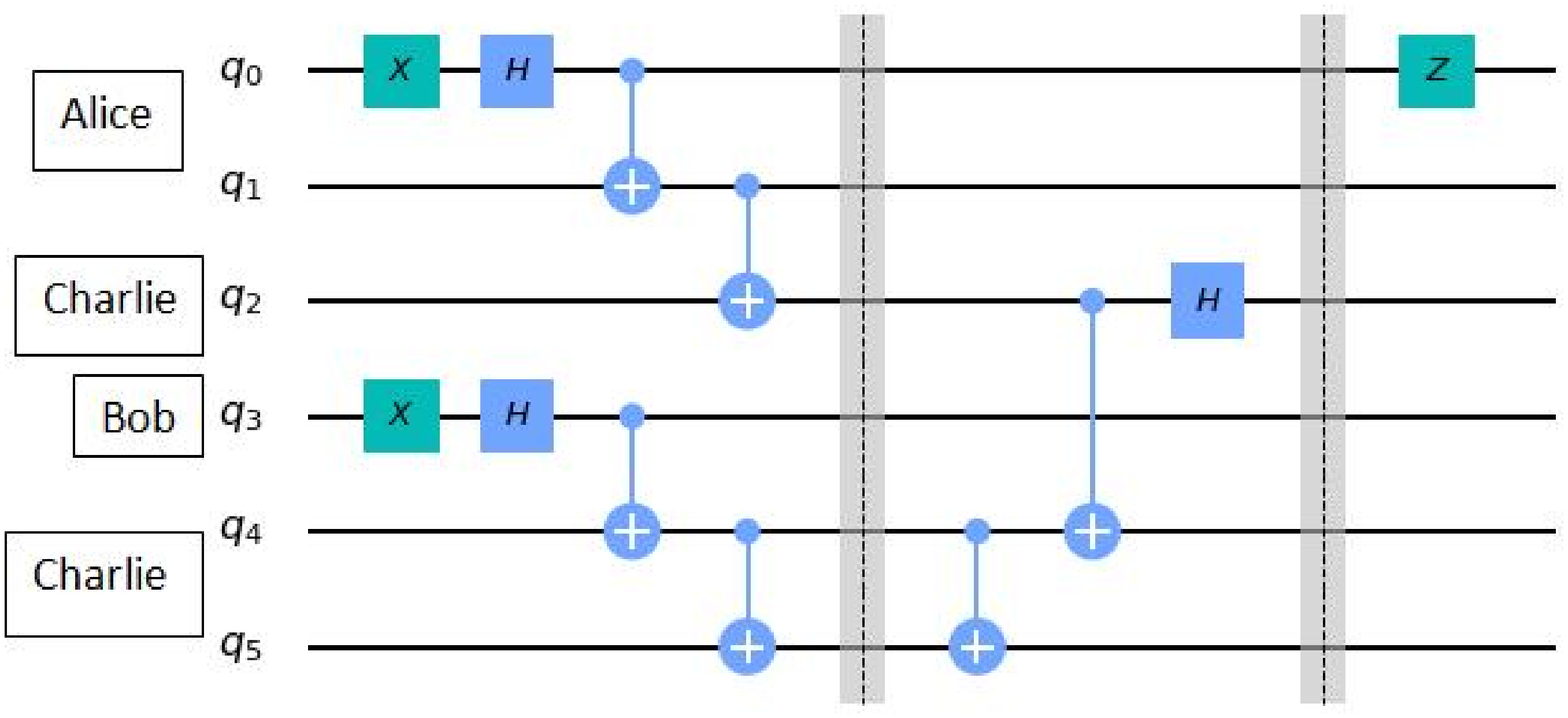}
\caption{Circuit implementation for entangled state}
\label{qsdc_Fig18}
\end{figure}

\subsection{Circuit implementation for Superdense coding}
We assume that after GHZ basis measurement, due to entanglement swapping the entangled state between Alice and Bob is the GHZ state $\ket{\psi_{000}}_{A_{1}A_{2}B}$. We perform superdense coding operations for three qubit states on Alice's two qubits in $M_A$ and randomly apply Z gate on Bob's single qubits in $M_B$. $M_A$ and $M_B$ are now sent to Charlie, the third party, who now performs GHZ basis measurement and publishes the results. We implement a circuit to demonstrate the same.

\begin{figure}[H]
\centering
\includegraphics[scale=0.3]{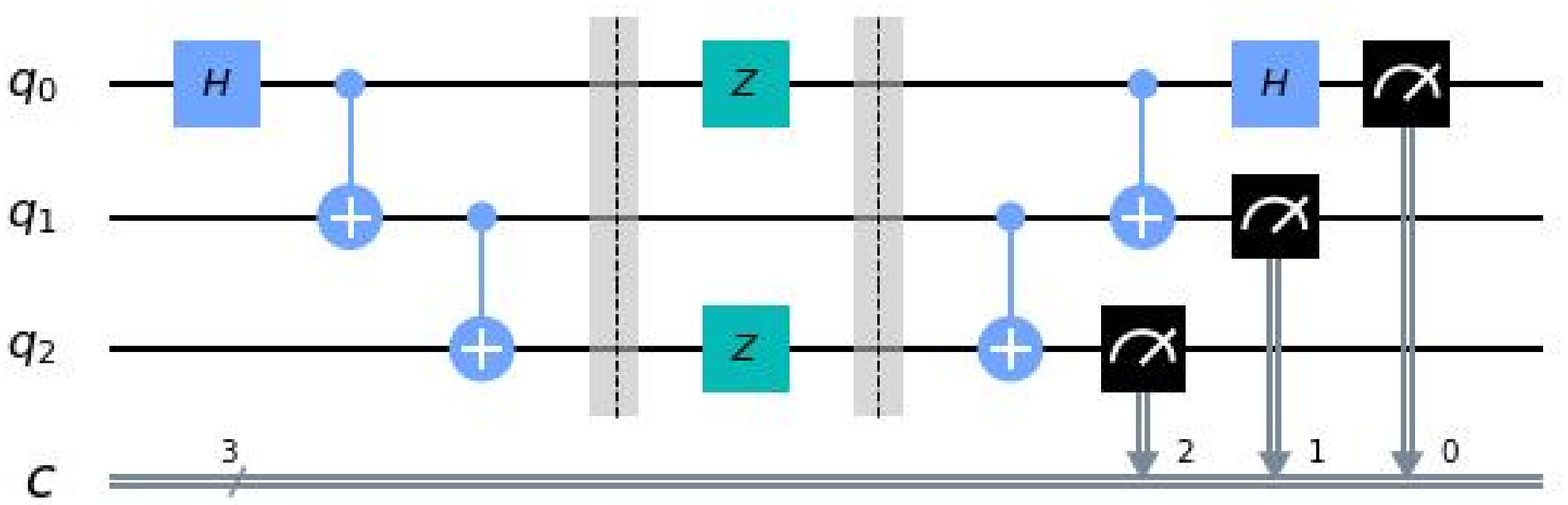}
\caption{Superdense coding circuit}
\label{qsdc_Fig19}
\end{figure}

After plotting the histogram, we again find the message received by Bob. Bob can then decode the message, as it is known to him through superdense coding operations what state Bob is supposed to receive before and after applying the Z gate.\\

\begin{figure}[H]
\centering
\includegraphics[scale=0.3]{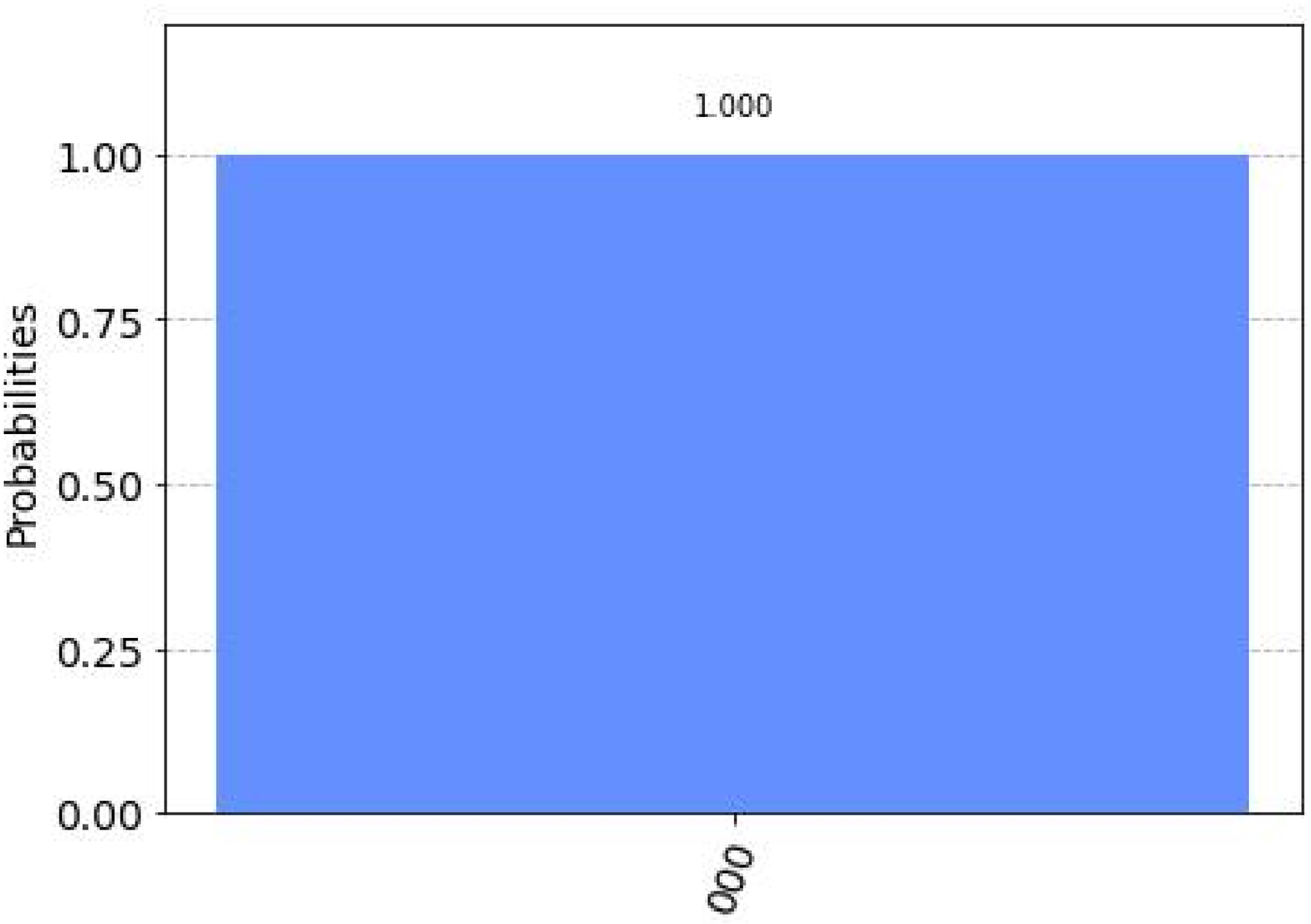}
\caption{Histogram plot of the message received by Bob}
\label{qsdc_Fig20}
\end{figure}

Similarly, we can check for all other superdense coding operations. Since, it is only known to Bob where he has applied Z gates, Charlie cannot know the secret message sent by Alice to Bob. After the results are published by Charlie, Bob compares it with Alice to decode the message.

\section{Discussion \label{qsdc_Sec4}}
Here, we have realized the MDI-QSDC protocol by implementing quantum circuits for Bell states in Sec. \ref{qsdc_Sec2} and for GHZ states in Sec.\ref{qsdc_Sec3}. We make all the necessary calculations required for security check and decide whether the communication process should proceed or not. This can be generalized for other maximally entangled states and quantum circuits and the necessary protocols can be implemented for those. The MDI-QSDC protocol solves the problem of defects in practical measurement devices by eliminating the loopholes in the  measurement device while making the communication process practically immune to interception by an eavesdropper. This improves the security of communication between two parties which is vital in modern-day communication systems.

\section*{Acknowledgments}
\label{fpi_acknowledgments}
A.G. would like to thank Bikash's Quantum (OPC) Pvt. Ltd. and IISER Kolkata for providing hospitality during the project work. B.K.B. acknowledges the prestigious Prime Minister's Research Fellowship awarded by DST, Govt. of India. The authors acknowledge the support of IBM Quantum Experience. The views expressed are those of the authors and do not reflect the official policy or position of IBM or the IBM Quantum Experience team.


\begin{IEEEbiographynophoto}{Arunaday Gupta}
received the M.Sc degree in Physics from the University of Delhi, Delhi, India, in 2019. He has undergone research internship at IISER Kolkata, West Bengal, India under the supervision of Prof. Prasanta K. Panigrahi and Mr. Bikash K. Behera from February to June 2020. His current research interests include quantum information theory, quantum communication, quantum cryptography, quantum simulation, quantum foundations and quantum thermodynamics.
\end{IEEEbiographynophoto}

\begin{IEEEbiographynophoto}{Bikash K. Behera}
is a PhD research scholar at the Department of Physical Sciences (DPS), IISER Kolkata, West Bengal, India. He is a recipient of the prestigious Prime Minister's Research Fellowship (PMRF) awarded by the Department of Science and Technology (DST), Govt. of India. He is also the Founder and CEO at Bikash's Quantum (OPC) Pvt. Ltd., Mohanpur, West Bengal, India. His areas of research interest are quantum information theory, quantum communication, quantum cryptography, quantum biology, quantum metrology and quantum game theory.
\end{IEEEbiographynophoto}

\begin{IEEEbiographynophoto}{Prasanta K. Panigrahi}
obtained his Ph.D. in quantum field theory from the University of Rochester, New York, USA. Currently he is a Professor at the Department of Physical Sciences (DPS), IISER Kolkata, West Bengal, India. His areas of research interest are field theory, quantum computing, quantum information processing, nonlinear dynamics, complex systems, wavelet analysis, computational physics and biomedical signal processing.
\end{IEEEbiographynophoto}
\end{document}